\documentclass[12pt,a4paper]{article}

\usepackage{amsmath, amssymb, amsfonts}
\usepackage{bm}
\usepackage{physics}
\usepackage{geometry}
\usepackage{graphicx}
\usepackage{hyperref}
\usepackage{enumitem}
\newcommand{\ber}{\begin{eqnarray}}
\newcommand{\eer}{\end{eqnarray}}
\newcommand{\bea}{\begin{equation}}
\newcommand{\eea}{\end{equation}}
\newcommand{\del}{\partial}
\begin{document}

\title{Dynamical Fermionization and Emergent Bethe Rapidity Structure in the Spatial Density of Cold quenched Lieb-Liniger gas}
\author{\bf Sumita Datta $^{1,2,a}$\\
{\small $^{1}$Department of Pure and Applied Mathematics, Alliance University,}\\
{\small Bengaluru 562 106, India}\\
{\small $^{2}$Department of Physics, University of Texas at Arlington,}\\
{\small Texas 76019, USA}\\
{\small $^a$ {\bf Corresponding Author:} sumitad@gmail.com}\\
\bf James M Rejcek $^{2,c}$\\
\bf Rajasee Datta $^{d}$\\
{\small Trinity College Dublin, The University of Dublin}\\
\bf Maxim Olshanii $^{b}$ \\
{\small Department of Physics, University of Massachusetts Boston}\\
{\small Boston MA 02125, USA}\\
{\small $^b$ maxim.olchanyi@umb.edu}\\
{\small $^c$ james.rejcek@sbcglobal.net}\\
{\small $^d$ dattar@tcd.ie}}

\date{\today}

\maketitle

\begin{abstract}
We demonstrate that the nonequilibrium spatial density of a one-dimensional interacting Bose gas, following a geometric quench, directly encodes information about the underlying momentum (rapidity) distribution of the system. 

Starting from the interacting ground state of a Lieb--Liniger gas confined in a hard-wall box of length $L_0$, we study its expansion into a larger box of length $L > L_0$ at fixed interaction strength. Using an ab initio quantum Monte Carlo approach based on the generalized Feynman--Kac representation, we compute the time evolution of the many-body density.

We show that, in the long-time limit, the density profile acquires a scaling form in the velocity variable $x/t$, approaching a stationary distribution whose shape reflects the underlying rapidity structure. The velocity-space density broadens systematically with increasing interaction strength and exhibits rapid convergence in the strongly interacting (Tonks--Girardeau) regime.

These results provide numerical evidence that ballistic expansion enables a direct mapping between spatial density profiles and the momentum-space structure of the integrable Lieb--Liniger model, offering a practical route to accessing Bethe rapidities through real-space observables.
\end{abstract}

\newpage

\section{Introduction}
In the history of physics, fields or particles initially considered to play auxiliary roles are often later recognized as observable physical phenomena. A prime example is the Aharonov-Bohm effect, which elevates the magnetic vector potential from a mere mathematical tool to an independent empirical entity. Similarly, the Bethe Ansatz \cite{1}, a powerful method for solving any-body and many-spin problems, involves mapping to free Gaudin fermions. These Gaudin fermions \cite{2}, represented by Bethe Rapidities (BRs), have become experimentally accessible only recently in studies on "Dynamical Fermionization" (DF) conducted at Pennsylvania State University \cite{3}.

The foundation of this research traces back to Marvin Girardeau's seminal work \cite{4} in 1960 on the correspondence between hard-core bosons and non-interacting fermions.
The mapping of bosonic wave functions to the absolute value of fermionic wave functions has profound consequences for the dynamical properties of
the system. In particular, when a strongly interacting Bose gas is
released from a confining potential, its expansion dynamics exhibit
fermionic features, even though the system remains bosonic at the
level of statistics. This phenomenon, known as \emph{dynamical
fermionization}, has been extensively studied both theoretically and
experimentally. This was further expanded by Yukalov and Girardeau \cite{35} in 2005, leading to the prediction of DF by Marcos Rigol et al[5]. in the same year. The experimental verification of DF by the AMO group at Pennsylvania State University in 2020 marked a significant milestone. This phenomenon is a frontier in the study of "strongly correlated systems" in condensed matter physics.

Motivated by these considerations, we investigate the nonequilibrium
dynamics of a Lieb--Liniger \cite{6} Bose gas released from a hard-wall box
into a larger box, corresponding to a geometric trap quench. Using an
\emph{ab initio} quantum Monte Carlo approach based on the generalized
Feynman--Kac path integral formulation \cite{9,10,11,12,22,23,24,25,26,33} we compute
the time evolution of the many-body spatial density across different
interaction regimes.

The integrability of the Lieb--Liniger model introduces an additional
layer of structure through the presence of conserved quantities known
as Bethe rapidities. These rapidities characterize the many-body
eigenstates and play a central role in determining the nonequilibrium
dynamics of the system. In the context of expansion dynamics, it has
been argued that the asymptotic spatial density profile, when expressed
in terms of the velocity variable $x/t$, becomes closely related to the
underlying rapidity distribution~\cite{7,20,32,34}.

This connection between spatial density and rapidity distribution is
reminiscent of other quantum phenomena where phase or momentum
information manifests indirectly in observable quantities as in the case of Aharonov-Bohm effect or Gaudin Fermions  mentioned earlier.
In the present context the rapidities are not directly measured, but their influence is encoded
in the emergent structure of the density profiles during ballistic
expansion.

Our primary objective is to examine how ballistic expansion reveals
the underlying integrable structure of the system, and in particular,
how the spatial density evolution encodes signatures consistent with
Bethe rapidities and dynamical fermionization. By analyzing the density
profiles in both real space and velocity space, we demonstrate the
emergence of stationary structures in the variable $x/t$ and their
dependence on interaction strength.

To our knowledge, DF and BRs have never been treated numerically from an \textit{ab initio} quantum perspective, particularly with QMC. Existing analytic solutions \cite{13, 14, 27, 28, 29} for a large class of many body and many spin problems using the Bethe Ansatz mostly cover the TG limit (interaction strength approaching infinity) with fewer particles.
Amongst other previous work on 1D gases, the expansion of 1D gases from power law traps using stationary phase and local density approximations \cite{31} and using Generalized Hydrodynamics\cite{15,16}, a strongly correlated 1D bose gas in a box trap and a harmonic trap \cite{30} are to name a few. Out of those, work done in \cite{31} and \cite{15,16} looks
promising.

The paper is organized as follows. In Sec.~2 we describe the model,
the quench protocol, and the numerical method based on the
Feynman--Kac representation. The results of the simulations are
presented in Sec.~3, where we analyze the spatial and velocity-space
density profiles and discuss their physical interpretation. Finally,
Sec.~4 contains our conclusions and outlook.

\section{Theoretical Basis for the Calculation of Numerical Evaluation of Initial Density, Asymptotic Density, and Bethe Rapidities}
\subsection{The Model and the Quench Protocol}

Our numerical description of DF and BRs is based on the important correspondence between impenetrable bosons and non-interacting fermions, established by Marvin Girardeau[4]. This impenetrability(
$\psi(x_1......x_n)=0$ if $x_j=x_l$, $1\le j<l\le n$;
where $x_1......x_n$ are the coordinates of n particles in the system) implies that for such systems, $\psi_0^B=|\psi_0^F|$, a concept known as fermionization.

When a TG gas is released from a trap (either a box or harmonic), its momentum and density distribution will asymptotically approach that of a spinless Fermi gas in the original trap, a phenomenon known as `Dynamical Fermionization'. The initial density can be evaluated using the ground state solution as given by Eq (11), $\rho_i = |\psi_i(x, t)|^2$, where $\psi_i(x, t)$ is the ground state solution of the quantum gas. To obtain the asymptotic solution after ballistic expansion, we consider the time evolution of the Hamiltonian and the solution in the long time limit. Using the propagator, the many-body solution can be written as

\begin{equation}
\psi(x, t) = \int k_p(x, y, t)\psi(y)dy
\end{equation}
where $k_p(x, y, t)$ is the time-dependent propagator. To calculate the asymptotic density, we use the long-time limit of this propagator, leading to $\rho_\infty = |\psi_\infty(x, t)|^2$. By quenching the initial Hamiltonian, i.e., changing the box length from $L_0$ to $L$ and allowing the gas to expand freely in the bigger box, theoretical predictions \cite{5,18,19} indicate that the initial real space density should resemble the asymptotic density. It has been analytically proven that $\rho_i = |\psi_i(x, t)|^2 \equiv \rho_\infty = |\psi_\infty(x, t)|^2 = |\psi_{F-NI}|^2$, where $ |\psi_{F-NI}|^2$ is the density of the non-intercating Fermi gas. Our interest is to observe that the density profiles during the expansion contain information about the initial state, demonstrating `Dynamical Fermionization'.

To observe the Bethe rapidities, the idea is to perform a ballistic expansion from the ground state of the box length $L_0$ to a free space in the interacting state. We propose to study the time evolution of the ground state in the long time limit. After expanding to a cloud size much greater than $L_0$, the spatial density will be proportional to the `Bethe Rapidities' \cite{7,20}.
We consider a one-dimensional system of $N$ bosons
interacting via a repulsive contact potential and
confined by hard-wall boundary conditions.
The initial Hamiltonian is
\begin{equation}
H_0 =
-\frac{\hbar^2}{2m}\sum_{i=1}^{N}\frac{\partial^2}{\partial x_i^2}
+ g \sum_{i<j}\delta(x_i-x_j)
+ \sum_{i=1}^{N} V_{L_0}(x_i),
\end{equation}
where $g>0$ is the interaction strength and
$V_{L_0}(x)$ denotes the hard-wall box potential of length $L_0$,
\begin{equation}
V_{L_0}(x) =
\begin{cases}
0, & 0<x<L_0, \\
\infty, & \text{otherwise}.
\end{cases}
\end{equation}

The system is initially prepared in the ground state
$\psi_0$ of $H_0$.

At time $t=0$, a sudden trap quench is performed.
The confining potential is instantaneously changed
from a box of length $L_0$ to a box of larger length $L>L_0$.
\begin{equation}
H =
-\frac{\hbar^2}{2m}\sum_{i=1}^{N}\frac{\partial^2}{\partial x_i^2}
+ g \sum_{i<j}\delta(x_i-x_j)
+ \sum_{i=1}^{N} V_{L}(x_i),
\end{equation}
with
\begin{equation}
V_{L}(x) =
\begin{cases}
0, & 0<x<L, \\
\infty, & \text{otherwise}.
\end{cases}
\end{equation}
Thus, the quench consists solely of a sudden
change in the trapping potential (geometric quench),
while the interaction strength remains unchanged.
The subsequent time evolution of the system
is governed by the post-quench Hamiltonian $H$,
\begin{equation}
\psi(x_1,\ldots,x_N;t)
=
e^{-iHt/\hbar}\psi_0(x_1,\ldots,x_N).
\end{equation}
\vspace{-30.0 mm}
\begin{figure}[h!]
\centering
\includegraphics[width=4.5in,angle=-0]{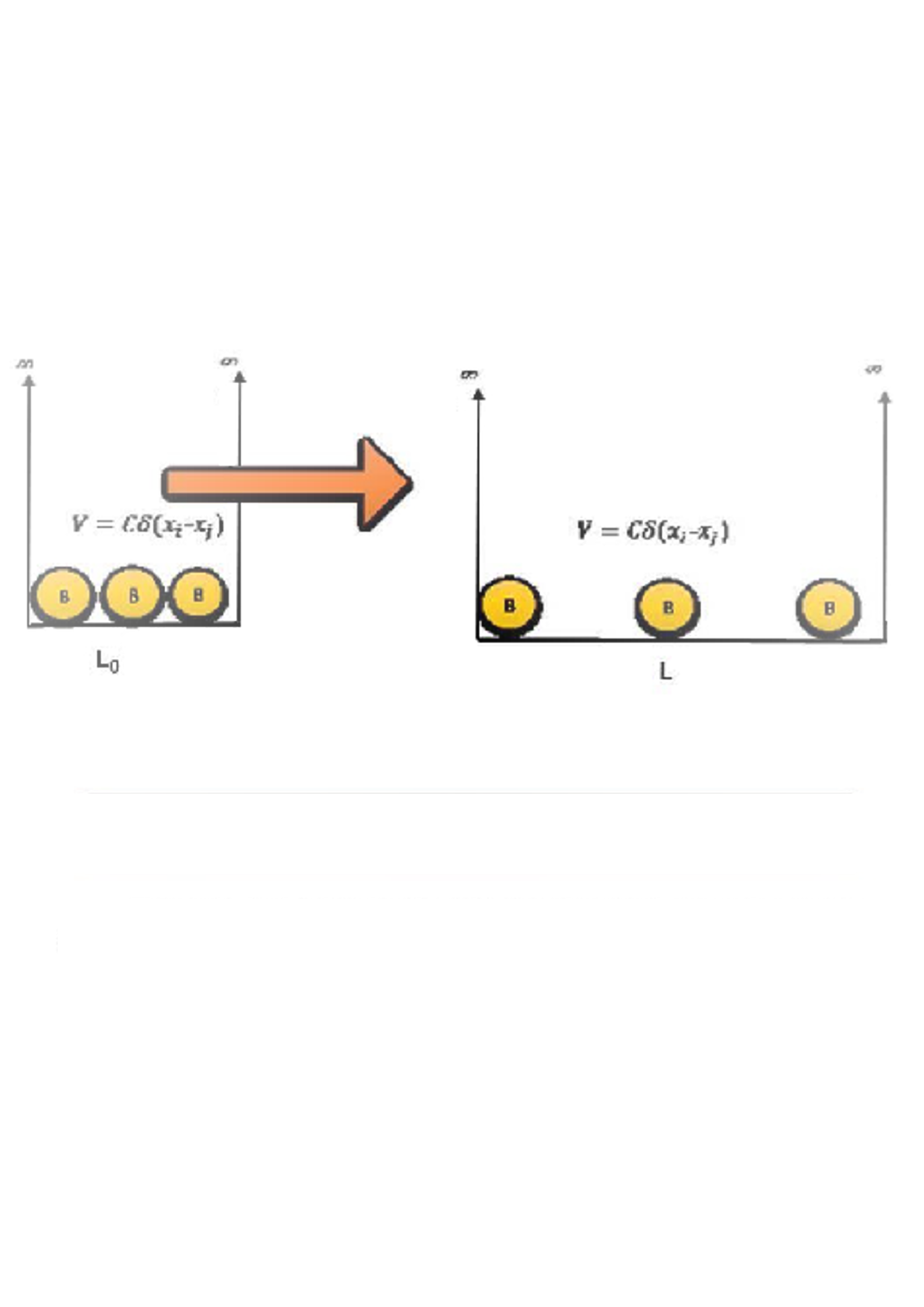}
\vspace{-64.5 mm}
\caption{A plot for the thought experiment of expansion of the gas from the box of Length $L_0$ to $ L$}
\end{figure}

\newpage
\subsubsection{Density Dynamics}
The many-body ground state and time evolution are
computed using a quantum Monte Carlo technique based
on the generalized Feynman–Kac path integral
representation.

In imaginary time the Schrödinger equation can be
written as

\begin{equation}
\frac{\partial \psi}{\partial t}
=
\left(
\frac{\Delta}{2}
-
V
\right)
\psi .
\end{equation}
The solution admits the stochastic representation

\begin{equation}
\psi(x,t)
=
E_x
\left[
e^{-\int_0^t V(X(s))ds}
f(X(t))
\right],
\end{equation}
here $X(t)$ denotes a diffusion process.
This representation allows the ground state and time-dependent wavefunctions to be obtained
numerically by sampling stochastic trajectories. To speed up the convergence one can formulate the generalized Feynman-Kac method we first rewrite the Hamiltonian as $H=H_0+V_p$,
where $ H_0=-\frac{\Delta}{2}+{\lambda}_T+\frac{{\Delta}{\phi}_T}{2{\phi}_T}$ and $V_p=V-({\lambda}_T+\frac{{\Delta}{\phi}_T}{2{\phi}_T})$. We choose the non-negative trial function to be a trial function associated with the symmetry of the problem i.e.,$\phi={\phi}_T$ and ${\lambda}_T$ is the trial energy of this reference function.
Then the new stochastic solution can be written in terms of the reference potential $V_p$ as follows:
\begin{equation}
\nu(x,t)=E_{x}[e^{-\int_{0}^{t}{V_p}(Y(s))ds}]
\end{equation}

The normalized version of the many body density in the coordinate represetation can be represented as
\bea
\rho(x_1,x_2,.......x_N;x_1^{\prime},x_2^{\prime},.......,x_N^{\prime}) \\
=\sum_i w_i{\psi}^{*}(x_1,x_2,.....x_N)\psi(x_1^{\prime},x_2^{\prime},.......,x_N^{\prime})
\eea
where $ w_i=\frac{ e^{-\beta E_i}}{\sum_i \beta E_i}$ \\
Here $\beta$ represents imaginary-time propagation and formally
corresponds to the inverse temperature in the canonical ensemble.
In the present work, imaginary-time evolution is employed
as a projection method to obtain the interacting ground state.
\bea
\rho(x_1,x_2,.......x_N;x_1^{\prime},x_2^{\prime},.......,x_N^{\prime},\beta) \nonumber \\
=\frac{\sum_i e^{-\beta E_i}{\psi}^{*}(x_1,x_2,.....x_N)\psi(x_1^{\prime},x_2^{\prime},.......,x_N^{\prime})}{\sum_i e^{-\beta E_i}}
\eea
where $\{\psi_i\}$ form a complete orthonormal set of
many-body eigenstates of $\hat{H}$ satisfying
$\hat{H}\psi_i = E_i\psi_i$.
Assuming that the ground state is non-degenerate
and that $E_0 < E_1$, the limit $\beta \rightarrow \infty$
suppresses all excited-state contributions.
For $\beta\rightarrow \infty $, only the ground state contributes or in other words,\\
\ber
\sum_i e^{-\beta E_i}{\psi}^{*}(x_1,x_2,.....x_N)\psi(x_1^{\prime},x_2^{\prime},.......,x_N^{\prime}) \nonumber \\
\rightarrow e^{-\beta E_0}{\psi}^{*}(x_1,x_2,.....x_N)\psi(x_1^{\prime},x_2^{\prime},.......,x_N^{\prime})
\eer
and
\bea
\sum_i e^{-\beta E_i }\rightarrow e^{-\beta E_0} .
\eea

Then the many body density becomes
\ber
\rho(x_1,x_2,.......x_N;x_1^{\prime},x_2^{\prime},.......,x_N^{\prime})
& & =\frac{e^{-\beta E_0}{\psi}^{*}(x_1,x_2,.....x_N)\psi(x_1^{\prime},x_2^{\prime},.......,x_N^{\prime})}{e^{-\beta E_0}} \nonumber \\
& &={\psi}^{*}(x_1,x_2,.....x_N)\psi(x_1^{\prime},x_2^{\prime},.......,x_N^{\prime})
\eer
Here we have assumed that the eigenfunctions are normalized,
$\int |\psi_0|^2 dx = 1$,
so that the resulting density matrix remains properly normalized.
In the  diagonal represetation density function becomes \cite{25} \\
$\rho(x_1,x_2,.......x_N;x_1,x_2,.......,x_N)=\psi_i^2(x_1,x_2,.......x_N) $.\\
In terms of Feynman-Kac solution the density becomes \\
\bea
\rho(x_1,x_2,.......x_N)=|E_{x}[e^{-\int_{0}^{t}V(X(s))ds}f(X(t))]|^2.
\eea
The normalized version of many body density can be provided as:
\bea
\rho(x_1,x_2,.......x_N)= {N_c} |E_{x}[e^{-\int_{0}^{t}V(X(s))ds}f(X(t))]|^2.
\eea
where $ N_c $ is the nomalization constant.
This is just the density matrix for the lowest energy state for $t=0$. For propagation time 't' the density matrix will be defined in terms of time dependent solution as follows:
\bea
\rho(x_1,x_2,......x_N;t)= {N_c} |\psi(x_1,x_2,x_3,......,x_N;t)|^2
\eea
where $\psi(x_1,x_2,x_3,......,x_N;t)=E_{x}[e^{-\int_{0}^{t}V(X(s))ds}f(X(t))]$. Note that
$\rho(x_1,x_2,.....,x_N;t)=f(x,g,t)$ which we plot in Fig 2-6.
\subsection{Correspondence between the asymptotic spatial density(our approach) and Bethe Rapidities of LL system}
In the Lieb-Liniger(LL) model, the Schr\"{o}dinger equation for a Bose gas with N particles in one dimension interacting with a repulsive $\delta$ function potential reads as \\
$ [-\sum_{i=1}^N\frac{\hbar^2}{2m}\frac{\del^2}{{\del x_i}^2}
+2c\sum_{i<j}\delta(x_i-x_j)]\psi=E\psi $\\
Using atomic units for $\hbar=2m=1$, the above equation takes the following form:\\
$ [-\sum_{i=1}^N\frac{\del^2}{{\del x_i}^2}
+2c\sum_{i<j}\delta(x_i-x_j)]\psi=E\psi $.
Also for repulsive interactions we have \\
$ c>0 $ and  $x\in R : \{x_i |0\le x_i \le L \}$.\\
For our present purpose, we consider the freely expanding $\delta$ interacting LL gas which was initially localized by a trapping potential(either box or harmonic) was represented as follows: 
\bea
H=\sum_{i=1}^N-\frac{\hbar^2}{2m}\frac{\del^2}{{\del x_i}^2}
+g_{1D}\sum_{i<j}\delta(x_i-x_j)+V_{trap}
\eea
In the TG limit, the strongly interacting bosons and free fermions share the same energy spectrum and their wave functions
are related as $\psi_0^B=|\psi_0^F|$ (see Ref[4]). This phenomenon is called Fermionization.\\
 In \cite{19} the expression  for the time-dependent solution for the above freely expanding LL gas is
reprsented as follows :\\
\ber
 {\psi}_{B,c}(x_1,x_2,......,x_N;t)=\int {dk_1......dk_N G(k_1,......k_N)e^{i\sum_{j=1}^N[k_jx_j-\omega(k_j)t]}}
\eer
Here the $x_js$ are the position coordinates, $k_js$ are the momentum coordinates and $\omega(k_j)=k_j^2$. Also t is the propagation time and subscript B stands for Bosons and c signifies the interaction strength. $G(k_1,......k_N)$ contains information about all the initial condition and related to projection co-efficients of the initial wavefunction onto hard wall Lieb-Liniger eigenstates.
The asymptotic wave functions take the following form:
\ber
{\psi}_\infty(x_1,x_2,......,x_N;t) \propto t^{-N/2} \prod_{1 \leq i < j \leq N}[sgn(x_i-x_j)+\frac{i}{c}(k_j^{\prime}-k_i^{\prime})] \\
\nonumber
\times {{\tilde {\psi}}_{F}(k_1^{\prime}\ldots k_N^ \prime)} e^{i\sum_{j=1}^N[k_j^{\prime}x_j-\omega(k_j^{\prime})t]}
\eer
Now the above equation can be expressed in spatial coordinates by substituting $k_j^{\prime}={x_j}m/t={x_j}/2t $:
\ber
{\psi}_\infty(x_1,x_2,......,x_N;t) \propto t^{-N/2} \prod_{1 \leq i < j \leq N}[sgn(x_i-x_j)+\frac{i}{2ct}(x_j-x_i)] \\ \nonumber
\times {{\tilde {\psi}}_{F}(x_1/{2t}\ldots x_N/{2t})} e^{i/{4t}\sum_{j=1}^N{{x_j}^2}}
\eer
In \cite{19} the single particle density matrix was defined as:
\ber
\rho_{B,c}(x,y,t)=N\int{dx_2......dx_N}{\psi}_{B,c}^{*}(x,x_2......x_N,t) {\psi}_{B,c}(y,x_2......x_N,t)
\eer
The correponding momentum distribution is defined as
\ber
n_B(k,t)=\frac{1}{2\pi}\int dx dy e^{ik(x-y)}\rho_{B,c}(x,y,t)
\eer
In Ref[19] it was also shown that the asymptotic form of the momentum distrubution of a freely expanding LL gas can be expressed  as:
\ber
n_{B,\infty}(k) \propto \int d{\xi}_2........d{\xi}_N|G(k,{{\xi}_2}/2....,....{{\xi}_N}/2)|^2
\eer
The corresponding single particle spatial density when expressed in convenient velocity variable $\xi=x/t $ takes the form
\ber
{\rho}_{\infty}(\xi) \propto \int d{\xi}_2........d{\xi}_N|G(\xi,{{\xi}_2}/2....,....{{\xi}_N}/2)|^2
\eer
Following Ref[41], one can show that the asymptotic single particle spatial density can be written as \\
\ber
{\rho}_{\infty}(x,t)=\frac{m}{\hbar t}g(\frac{mx}{\hbar t})
\eer
Similarly the corresponding momentum density takes the following form:
\ber
n_{\infty}(k,t)=g(k)
\eer
So using insights from \cite{18,19,31} it can be established that the aymptotic single-particle spatial density can be determined by rescaling the initial quasimomentum density $g(k)$ distribution and they share the same shape in the long time limit. We borrow this idea in the many body calculation associated with our thought experiment.
In our case we calculate the many body spatial density as given by Eq(13) and Eq(14). Although the rigorous derivation in \cite{18,19,31}
is formulated at the level of single-particle densities, we assume that the dominant scaling structure extends to the many-body density in the asymptotic ballistic regime.

\newpage
\subsection{Notation Table/Units}
\begin{table}[h!]
\begin{center}
\caption{\bf Notation Table}
\begin{tabular}{cl cl cl}
Notation/Phrase & Meaning \\
 $ c $                         & The strength $ \delta $ interaction in the LL model \\
 $  g_{1D}$                  & The strength of $ \delta $ interaction  in the path integral(PI) method \\
$ g(k)$                        & initial quasimomentum distribution function \\
$ \gamma $                    & The dimensionless parameter related to $c$ and $g_{1D}$ \\
$ \gamma=\frac{m g_{1D}N}{L}$ & The relation between $c$ and $ g_{1D}$ \\
$ \gamma=g_{1D}/2 $                & The relation between $ \gamma$ and $g_{1D}$ for  $N/L=1$ \\
$ f(k,\gamma)$                & The distribution function for the momentum density \\
$ f(x,g_{1D},t) $                  & The distribution function for the spatial density \\
$ \hbar=m=1$                  & The units used in path integral calculations \\
$ \hbar=2m=1 $                & The units chosen in LL model \\
$ x_{PI} $                    & The space co-ordinates in PI units \\
$ t_{PI} $                    & The time in PI units
\end{tabular}
\end{center}
\end{table}
subsection{Numerical details for the drifted random walk}
For guiding the random walk inside the box of length $L_0$ we use a trial function of the form:
\ber
\phi_0(x_1,x_2,x_3,.......,x_N)
& &=\phi_0(x_1)\phi_0(x_2).....\phi_0(x_N) \nonumber \\
& &=x_1(L-x_1)x_2(L-x_2).....x_N(L-x_N)
\eer
Using the drifted random walk and diffusion(Brownian random walk) inside a box of length $L_0$, we get the numerical solution for the ground state. The above trial function is used to generate the the drifted random walk. The diffusion part is generated as follows:

\subsection{Physical Regimes and Choice of Parameters}
The observation of ballistic scaling and the emergence of
velocity-space structures depend sensitively on the choice of
interaction strength, particle number, and propagation time.
In particular, resolving the crossover from weakly interacting
to Tonks–Girardeau regimes requires parameters that correctly
capture the relevant physical scales.

For this reason, the selection of numerical parameters is not
merely technical, but is essential for revealing the density
patterns discussed in the Results section.

For the best selection of the parameters, one needs to check the following criteria.
(i) The interactions should be truely two body in nature.
(ii) One needs to know the propagation time to determine whether one can expect to see Bethe rapidities. The expansion velocity time the
time must exceed the initial cloud size. Estimates for the expansion velocity depend on the regime we are at.
(iii) We must make sure that our choices for the coupling constants reflects the three principal regimes.
ideal gas vs mean-field vs Tonks-Girardeau.
For a box in these three regimes our parameters are bounded  by the following conditions respectively:\\
$ gN << \frac{{\hbar}^2}{mL^2}L$ \\
$\frac{{\hbar}^2}{mL^2}L << gN << \frac{{\hbar}^2}{mL^2} N^2 L$ \\
$\frac{{\hbar}^2}{mL^2}L << gN $
where $L$ is the length of the box, $N$ is the total number of the particles in the box and $m$ is the mass of the particles. We choose $\hbar=m=1$ in our units.
When we use Gaussian or any hill of height $u$ and width $\sigma$ we have to make sure that the negative of that is shallow enough to support only one bound state:$\frac{{\hbar}^2}{(m {\sigma}^2)}>> u $.
Also, we must make sure that the one particle de-Broglie wavelenghts involved are longer than that of $\sigma$: $\frac {L_{int}}{N} >> \sigma $ as Girardeau end was used
 to estimate the de-Broglie wavelength.
Number of partcles N=100 \\
Initial Length of the box $L_{int}=0.25$ \\
Final Length of the box $L_{fin}=1.0$  \\
$g_{1D}=162$, $g_{1D}=50$ \\
$m=1=\hbar=1$ \\
Our main analysis is related to the interaction strength $g_{1D}=162$,\\
$t_{Prop}=(0.000625,0.00078125,0.0015625,0.00234375,0.0025)$. The choice of the parameters can be justified
as follows:
One can think of a small parameter
$\epsilon =1/N$. Both the initial ($L_{int}$) and the final ($L_{fin}$) sizes of the box are of the order of 1; $\gamma$ is also (for a good half of the points) is of the order of 1.
Hence:
\begin{description}[font=$ \bullet $]
\item Density scales as $n \sim N/L \sim (1/\epsilon) (1/L)$.
\item 1D scattering length $a\sim \gamma 1/n \sim \epsilon L$.
\item $g\sim \frac{{\hbar}^2}{ma}\sim \frac{{\hbar}^2}{m}(1/L)(1/\epsilon) $.
\item Fermi Velocity scales as $ v_F\sim \hbar n/m \sim (\hbar/m)(1/L)(1/\epsilon)$.
\item Minimal propagation time to be able to convert the initial velocity distribution to the final distribution is $ t_{min}\sim L/v_F
        \sim \frac{{mL^2}\epsilon}{\hbar} $
\end{description}
These parameter choices ensure that the expansion reaches the
ballistic regime within the simulated time window, thereby
allowing the emergence of the velocity-space structures
observed in Figs.~2–7.
\section{Results and Discussion}

\subsection{Time Evolution of the Spatial Density}

We compute the time-dependent many-body spatial density
\begin{equation}
\rho(x,t) = N_c \, |\psi(x_1,\dots,x_N;t)|^2,
\end{equation}
where the initial state corresponds to the interacting ground state in a hard-wall box of length $L_0$.

Figures 2 and 3 show the real-space density evolution for interaction strengths $g=50$ and $g=162$, respectively, following the sudden expansion into a larger box of length $L$.

At $t=0$, the density corresponds to the confined ground state. As time increases, the cloud expands and the peak density decreases. For $g=50$, the profiles broaden continuously with time, whereas for $g=162$ the density profiles at later times become nearly coincident even in real space, indicating that the system has reached its asymptotic regime within the finite box.

However, real-space broadening alone does not determine the nature of transport. To distinguish between ballistic and diffusive dynamics, it is necessary to analyze the density in an appropriate scaling variable.

\subsection{Density in Velocity Representation}

To probe the scaling behavior, we introduce the velocity variable
\begin{equation}
v = \frac{x}{t}.
\end{equation}

The same densities are then plotted as functions of $x/t$ (Figs. 6 and 8). In this representation, the profiles corresponding to different times approach a common shape.

For $g=50$, the convergence toward a stationary curve is gradual, while for $g=162$ the overlap is already essentially complete at the largest propagation times. This indicates that the density acquires a time-independent structure in velocity space.

Such behavior is consistent with ballistic expansion, characterized by
\begin{equation}
\rho(x,t) \sim \frac{1}{t} F\!\left(\frac{x}{t}\right),
\end{equation}
where $F(v)$ is a stationary function. The observed collapse of the profiles in $x/t$ implies that the characteristic length scale grows linearly in time.

In contrast, diffusive dynamics would exhibit scaling in $x/\sqrt{t}$. The present results therefore demonstrate that the expansion is ballistic.

\subsection{Interaction Dependence and Ballistic Time Scale}

The approach to the asymptotic regime depends strongly on the interaction strength.

For intermediate interaction ($g=50$), the real-space density continues to evolve over the accessible time window, although the velocity-space representation shows clear convergence. For stronger interaction ($g=162$), both real-space and velocity-space densities become effectively stationary at large times.

This behavior can be understood in terms of an interaction-dependent velocity scale. In the strongly interacting regime, the system approaches the Tonks--Girardeau limit, where the dynamics are governed by an effective Fermi velocity $v_{\mathrm{eff}}$. The characteristic time required to reach the asymptotic regime can be estimated as
\begin{equation}
t_{\min} \sim \frac{L}{v_{\mathrm{eff}}}.
\end{equation}

As $g$ increases, $v_{\mathrm{eff}}$ increases and $t_{\min}$ decreases, explaining the faster convergence observed for $g=162$.

Because the expansion takes place in a finite box, once the accessible region is fully explored the density becomes stationary. The saturation observed in real space at large $g$ is therefore a finite-size effect superimposed on ballistic dynamics.
subsection{Numerical Evidence for Dynamical Fermionization}

Dynamical fermionization refers to the emergence of fermionic characteristics in an expanding one-dimensional Bose gas.

A key signature of this phenomenon is the appearance of a stationary velocity-space density of the form
\begin{equation}
\rho(x,t) \sim \frac{1}{t} F\!\left(\frac{x}{t}\right),
\end{equation}
where the function $F(v)$ reflects the underlying rapidity (quasimomentum) distribution.

Our numerical results exhibit this behavior across interaction regimes. In both cases studied, the density profiles approach a time-independent structure when expressed in terms of $x/t$, with faster convergence at larger $g$.

In the strongly interacting regime ($g=162$), the rapid stabilization and the broadened velocity distribution are consistent with the fermionization picture, where the dynamics resemble those of non-interacting fermions at the level of spatial observables.

Although we do not compute the momentum distribution explicitly, the emergence of a stationary velocity-space profile provides indirect evidence that the long-time density encodes the same structural information as the rapidity distribution.

\subsection{Physical Interpretation}

The central observation is the emergence of a time-independent density profile in the velocity variable. This indicates that the expansion reorganizes the initial state into a ballistic distribution characterized by a fixed velocity profile.

The interaction dependence of this profile reflects the underlying integrable structure of the Lieb--Liniger model. Increasing interaction strength leads to systematic broadening, consistent with an increasing effective Fermi scale.

Thus, the expansion dynamics reveal both the ballistic nature of transport and the interaction-dependent structure of the many-body state.

\subsection{Qualitative Comparison with Bethe--Ansatz Results and Area Conservation}

For completeness, we compare the velocity-space profiles with the rapidity distributions obtained from Bethe--Ansatz studies (Fig.~4).

This comparison is qualitative: no direct extraction of rapidities is performed. Nevertheless, the observed broadening of the velocity distribution with increasing interaction strength follows the same trend as the exact rapidity distributions.

An additional consistency check is provided by particle-number conservation:
\begin{equation}
\int \rho(x,t)\,dx = N.
\end{equation}
Numerically, the area under the density curves remains invariant across propagation times (Fig.~5), confirming that the observed evolution corresponds to redistribution rather than loss of particles.

\newpage
\begin{figure}[h!]
\includegraphics[width=5.5 in,angle=-0]{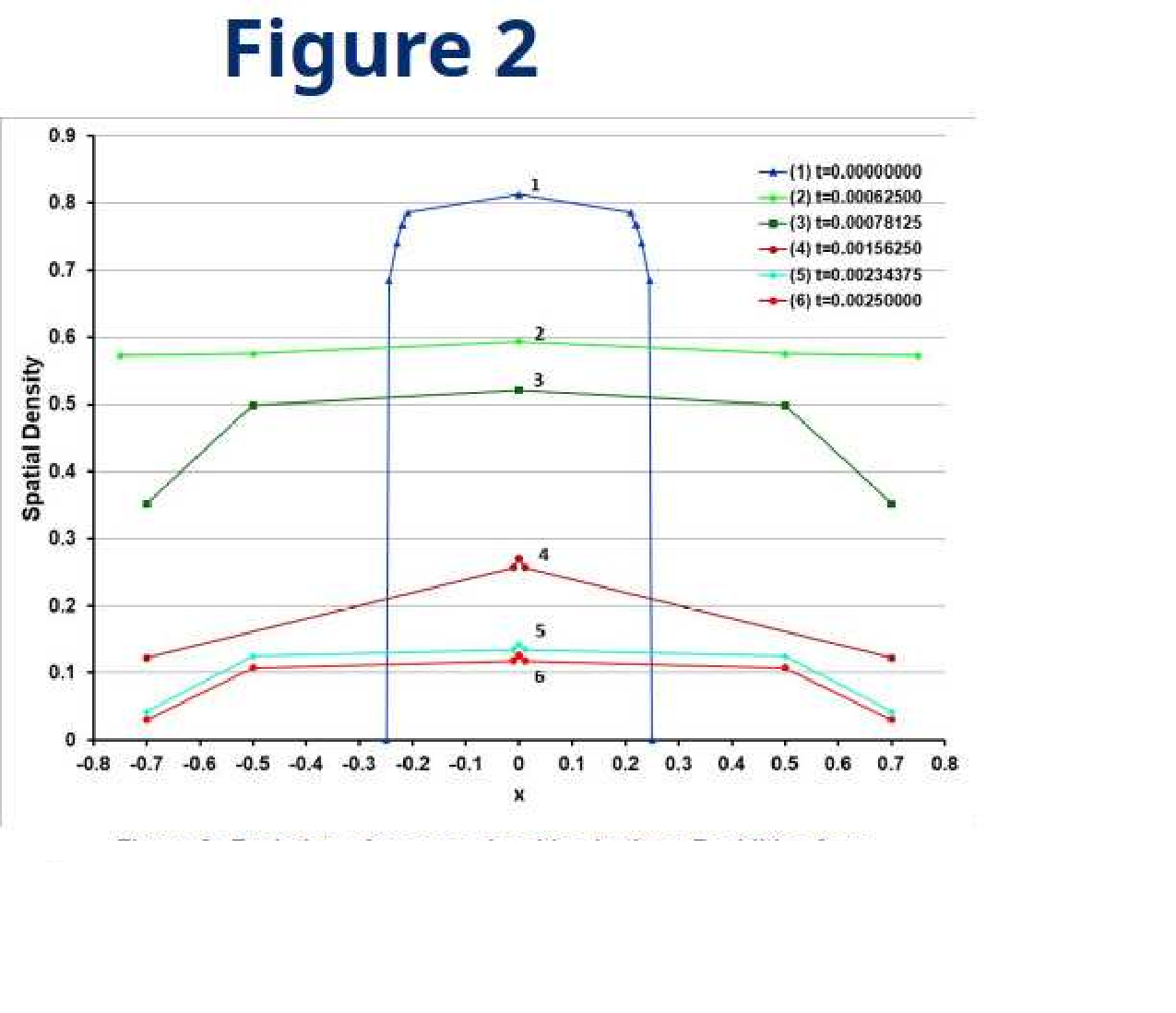} \\
{\bf Figure 2, Time evolution of the real-space density $\rho(x,t)$ for interaction strength $g=50$; Curve (1) corresponds to the initial ground state ($t=0$), while curves (2)–(6) represent densities at increasing propagation times $t = 0.000625,\ 0.000728,\ 0.0015625,\ 0.00234375,\ 0.0025$. The progressive broadening and reduction in peak height reflect the expansion of the gas. Number of particles: $N=100$.}
\end{figure}
\begin{figure}[h!]
\includegraphics[width=5.5 in,angle=-0]{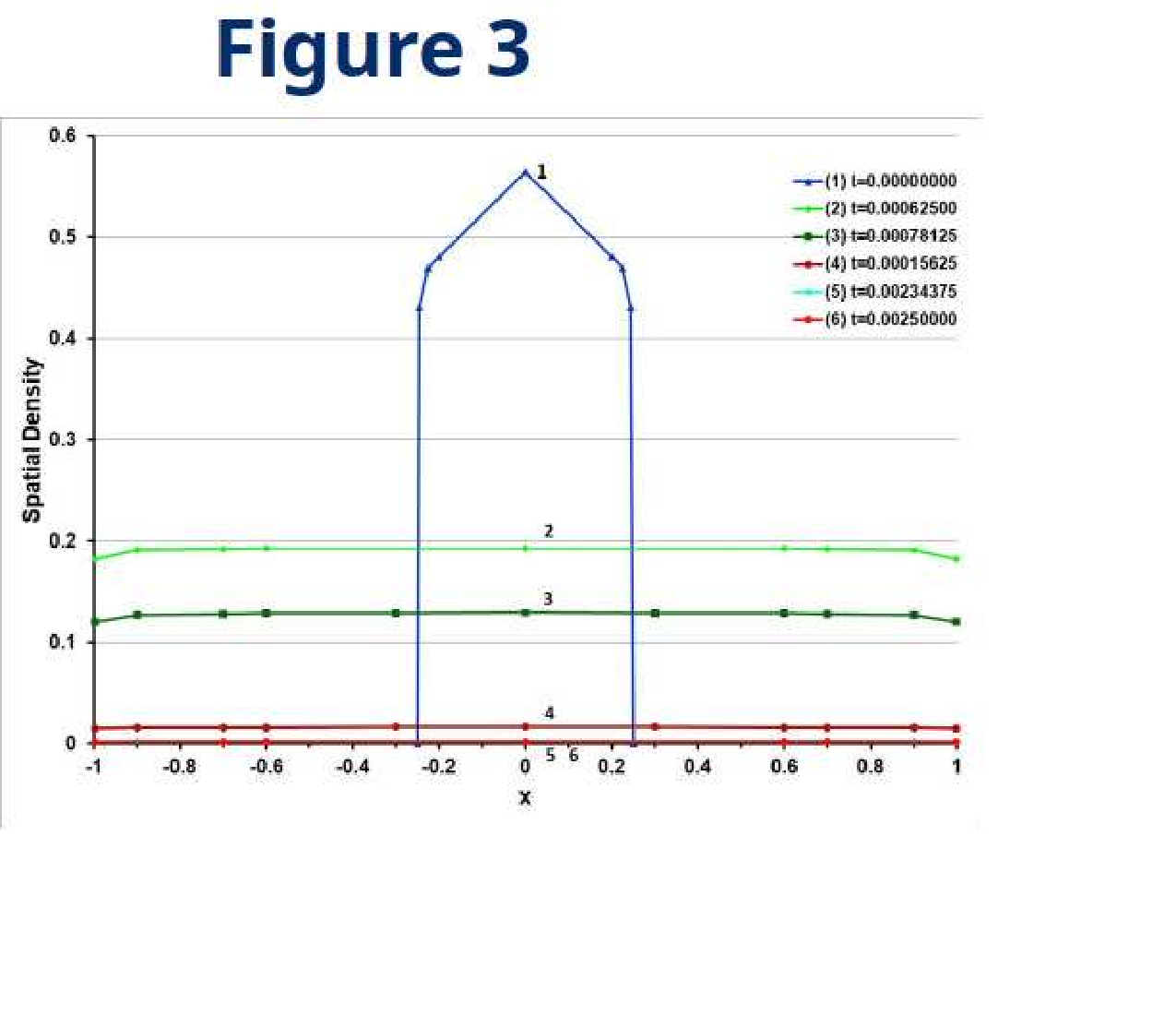} \\
{\bf Figure 3, Time evolution of the real-space density $\rho(x,t)$ for strong interaction $g=162$.
Curve (1) denotes the initial ground state ($t=0$), and curves (2)–(6) correspond to increasing propagation times
$t = 0.000625,\ 0.000728,\ 0.0015625,\ 0.00234375,\ 0.0025$.
At large times the profiles nearly overlap, indicating rapid convergence to the asymptotic regime.
Number of particles: $N=100$.}
\end{figure}
\begin{figure}[h!]
\includegraphics[width=7.5in,angle=0]{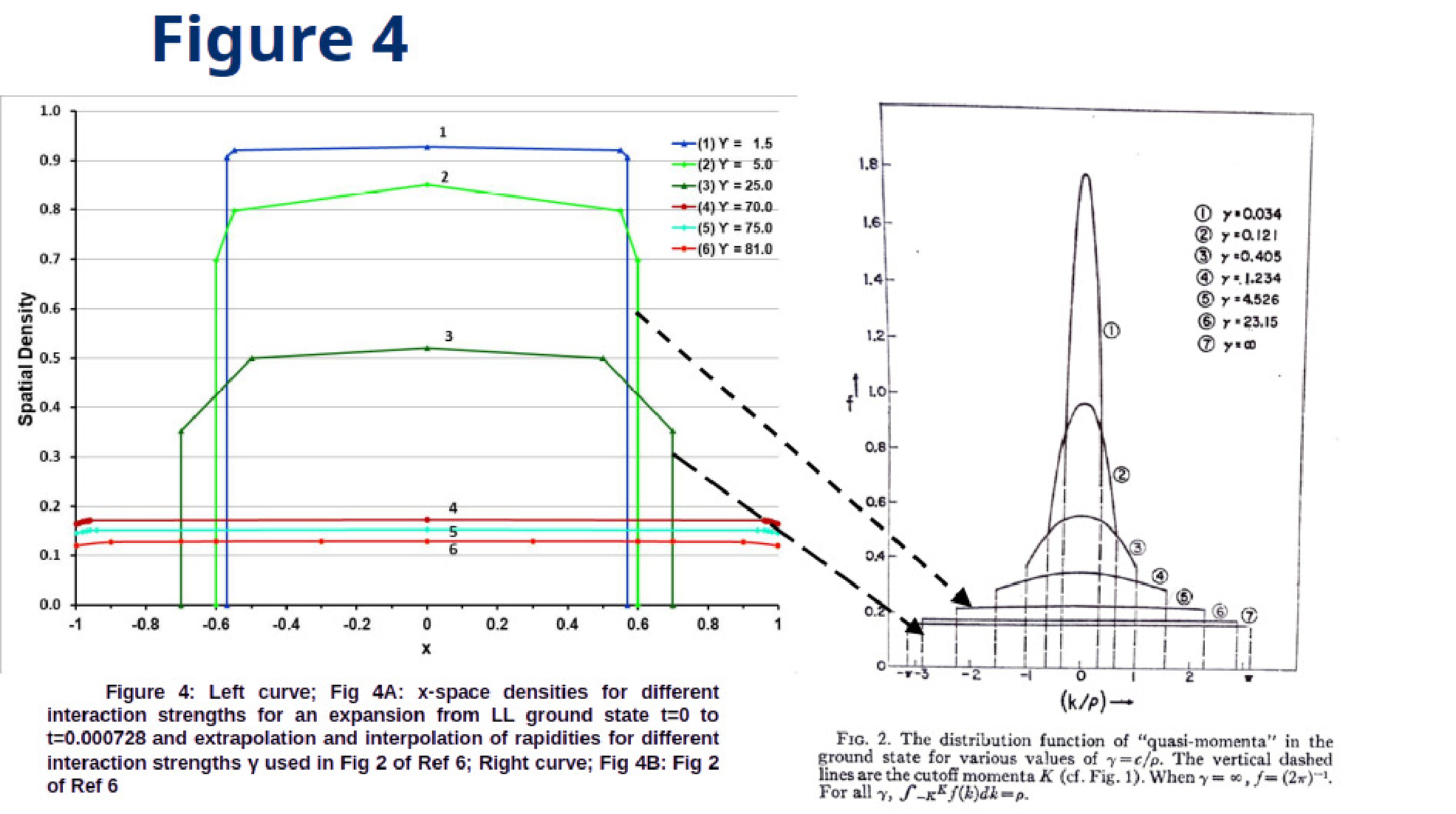}
\end{figure}
\begin{figure}[h!]
\includegraphics[width=7.5in,angle=-0]{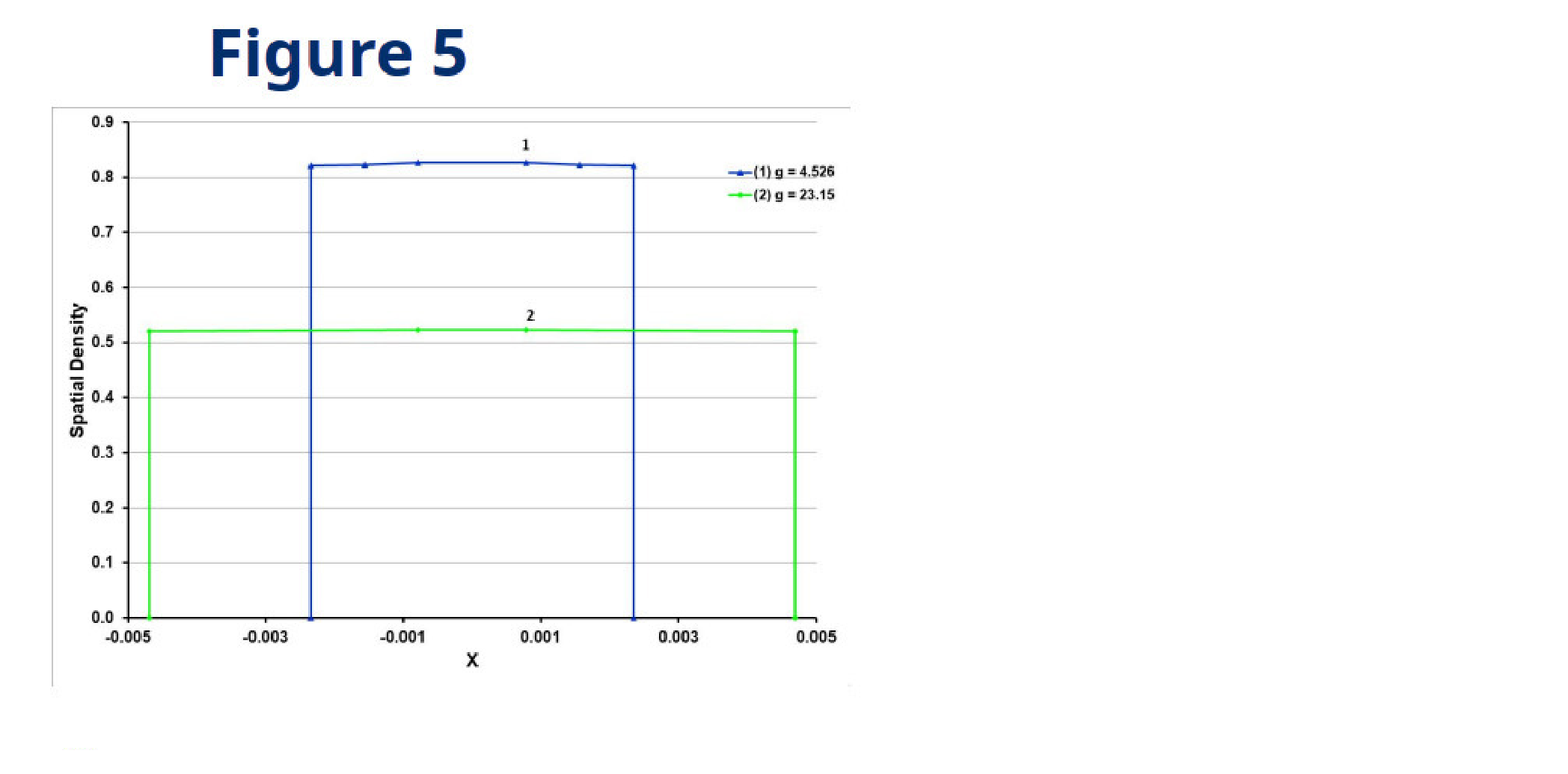}\\
{\bf Figure 5, Real-space density $\rho(x,t)$ for two interaction strengths at an early stage of expansion.
The comparison illustrates conservation of particle number through invariance of the area under the curves,
despite redistribution of density. Parameters correspond to $t=0$ and $t=0.000728$ with $N=100$.}
\end{figure}
\begin{figure}[h!]
\includegraphics[width=4.5 in,angle=-90]{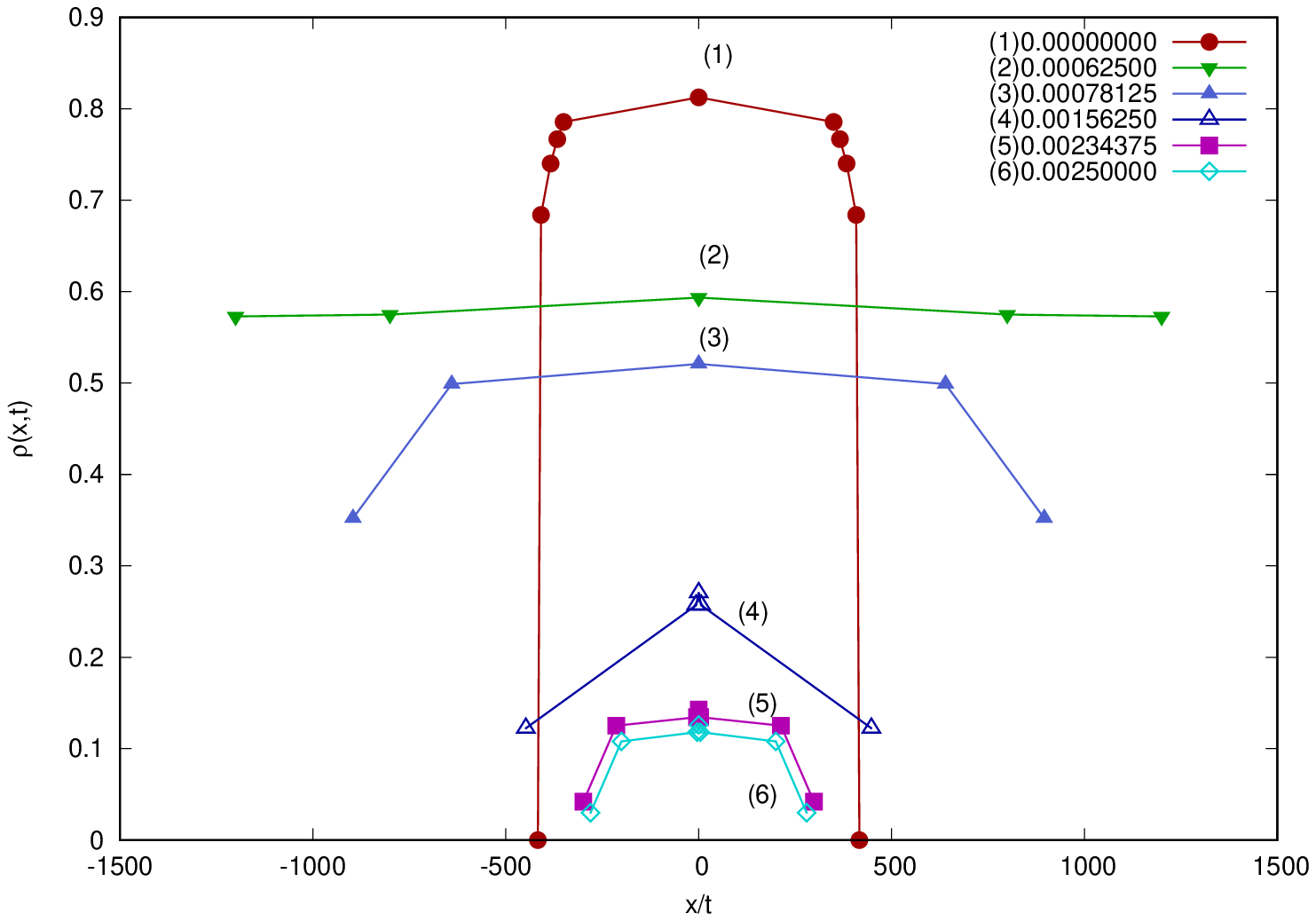} \\
{\bf Figure 6, Density profiles plotted in velocity space as $\rho(x,t)$ versus $x/t$ for interaction strength $g=50$.
Curve (1) represents the initial state ($t=0$), while curves (2)–(6) correspond to increasing times
$t = 0.000625,\ 0.000728,\ 0.0015625,\ 0.00234375,\ 0.0025$.
The gradual collapse of the curves indicates the emergence of a stationary velocity distribution,
consistent with ballistic expansion. Number of particles: $N=100$.}
\end{figure}
\begin{figure}[h!]
\includegraphics[width=4.5 in,angle=-90]{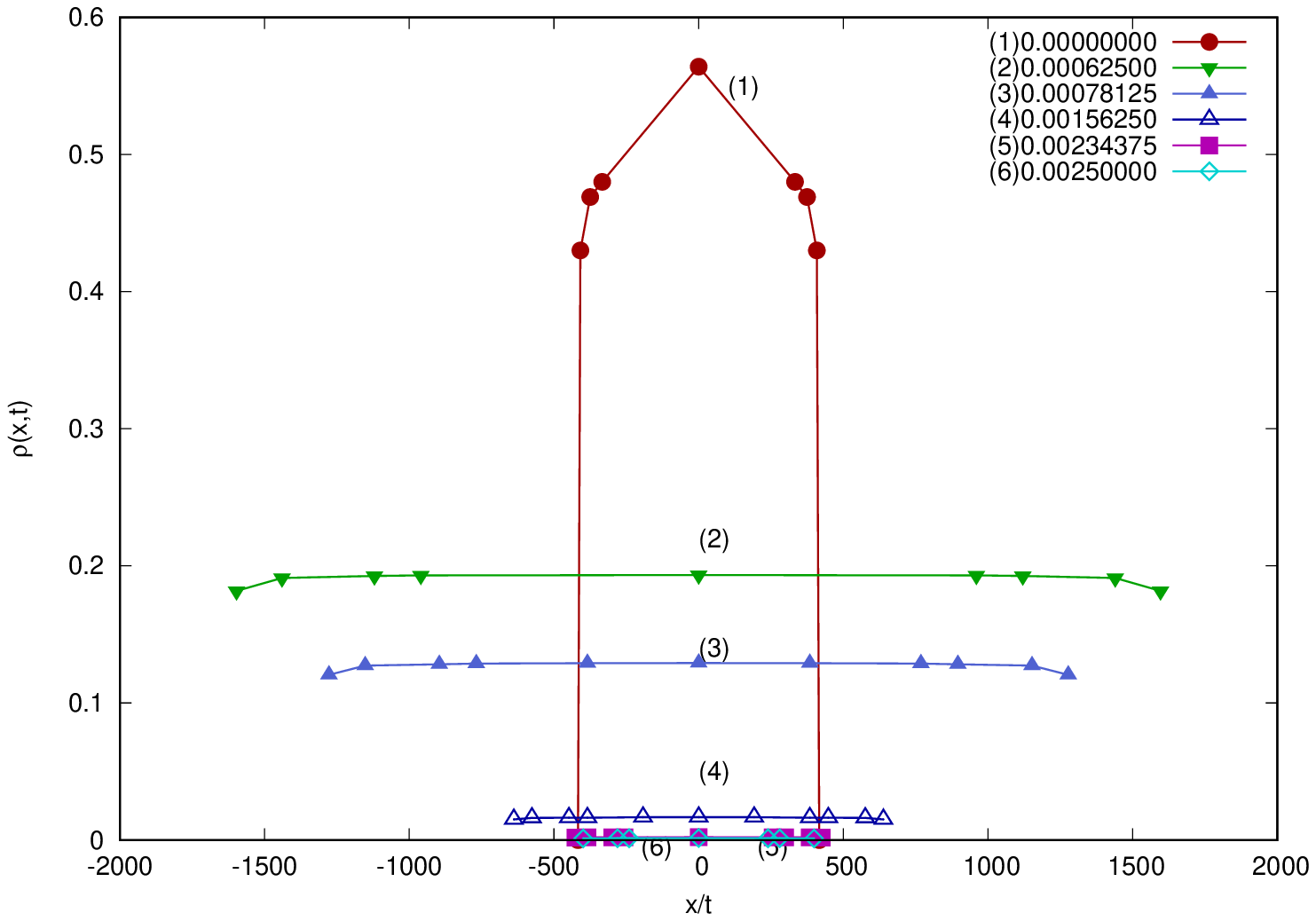} \\
{\bf Figure 7,
Velocity-space representation of the density, $\rho(x,t)$ versus $x/t$, for strong interaction $g=162$.
Curve (1) corresponds to the initial state ($t=0$), and curves (2)–(6) denote increasing propagation times
$t = 0.000625,\ 0.000728,\ 0.0015625,\ 0.00234375,\ 0.0025$.
The near-complete overlap of the profiles at large times demonstrates rapid convergence to a stationary
velocity distribution, characteristic of ballistic dynamics. Number of particles: $N=100$.}
\end{figure}
\clearpage
\subsection{Summary of Numerical Results}

The numerical results may be summarized by referring to
Figs.~2--7.

Figures~2 and~3 display the real-space evolution of the
many-body density following the trap quench for
interaction strengths $g=50$ and $g=162$, respectively.
In both cases, the density spreads into the larger box
and the peak height decreases with time.
For $g=50$, the profiles continue to broaden gradually,
whereas for $g=162$ the density profiles corresponding
to the largest propagation times become nearly coincident,
indicating that the system has already reached its
asymptotic regime within the finite box.

To clarify the nature of the expansion, the same data are
replotted in Figs.~6 and~8 as functions of the velocity
variable $v=x/t$ for $g=50$ and $g=162$, respectively.
In this representation, the density profiles approach
a stationary form as time increases.
The overlap becomes progressively clearer for $g=50$,
while for $g=162$ the velocity-space profiles are already
essentially time-independent at large propagation times.
This behavior demonstrates linear scaling of the
characteristic length with time and is consistent with
ballistic expansion.

Figure~4 presents a qualitative comparison of the
interaction-dependent velocity profiles with the
rapidity distributions obtained in the classical
Bethe--Ansatz solution of the Lieb--Liniger model.
Although no direct extraction of rapidities is performed,
the systematic broadening of the velocity distribution
with increasing interaction strength follows the same
trend as the exact rapidity distributions.
This agreement supports the interpretation that the
long-time spatial density reflects the underlying
Bethe rapidity structure.
inally, Fig.~5 illustrates conservation of the
total area under the density curves for a fixed
temperature.
Within numerical accuracy, the integral
\begin{equation}
\int \rho(x,t)\,dx
\end{equation}
remains constant throughout the evolution.
This confirms particle-number conservation and
demonstrates that the observed redistribution
of density is due to ballistic spreading rather
than numerical artifacts or loss of normalization.

The spatial density at two different intercation strengths $ g=50$ and $ g=162 $ are plotted in Fig.~ 6 and Fig.~ 8 respectively.

Taken together, Figs.~2--7 establish that the
post-quench dynamics are ballistic, interaction-dependent,
and consistent with the emergence of Bethe rapidities
and dynamical fermionization in the expanding
Lieb--Liniger gas.

The numerical results may be summarized as follows:

(i) The post-quench expansion is ballistic, as demonstrated by the collapse of density profiles in the variable $x/t$.

(ii) The asymptotic velocity-space distribution depends on the interaction strength and broadens with increasing $g$.

(iii) Strong interactions lead to faster convergence to the asymptotic regime and exhibit fermion-like features consistent with dynamical fermionization.

(iv) Qualitative agreement with Bethe--Ansatz trends and conservation of particle number further support the interpretation.

\section{Conclusions and Outlook}

We have investigated the nonequilibrium dynamics of a one-dimensional interacting Bose gas following a geometric quench using an ab initio quantum Monte Carlo approach.

The results demonstrate that the expansion is ballistic and that the density approaches a stationary form when expressed in the velocity variable $x/t$. The corresponding velocity-space distribution depends systematically on interaction strength and becomes broader in the strongly interacting regime.

In the Tonks--Girardeau limit, the rapid convergence and fermion-like structure of the density profiles are consistent with dynamical fermionization. The emergence of a stationary velocity distribution indicates that the long-time spatial density encodes information about the underlying rapidity distribution, although no direct extraction is performed.
This work provides a numerical realization of ballistic integrable dynamics in a finite system and complements existing analytical approaches by directly resolving the time evolution of the many-body density.

Future work may include explicit computation of momentum distributions, extension to larger system sizes, and investigation of interaction quenches. These directions may further clarify the role of integrability and finite-size effects in nonequilibrium dynamics of one-dimensional quantum gases.

\clearpage
\newpage

{{\bf Acknowledgements}: One of the authors(SD) would like to thank Alliance University for providing support for carrying out the research work}\\.

{\bf Declaration of interests:} All the authors have no conflicts of interest to declare. All four authors have seen and agreed with the contents of the manuscript and there is no financial interest to report.\\

{\bf Data availability statement:} No data in this publication is to be made available under the study-participant privacy protection clause.\\

\end{document}